\documentclass[a4paper]{article}
\usepackage{issp2020,amssymb,amsmath,graphicx}
\usepackage{filecontents}
\usepackage{tabularx}

\usepackage[utf8]{inputenc}
\usepackage{adjustbox}
\usepackage{subcaption}
\usepackage{multirow}
\usepackage{graphicx,subcaption,multicol}
\usepackage[T1]{fontenc}
\usepackage[english]{babel} 
\usepackage[autostyle]{csquotes}
\usepackage[backend=biber,style=authoryear,sorting=nyt]{biblatex}
\defbibheading{refs}{\section{References}}
\bibliography{ref} 

\ninept

\title{Inverted Vocal Tract Variables and Facial Action Units to Quantify Neuromotor Coordination in Schizophrenia}

\vspace{-15pt}

\makeatletter
\def\name#1{\gdef\@name{#1\\}}
\makeatother
\name{{ Yashish Maduwantha H.P.E.R.S$^1$, Chris Kitchen$^2$, Deanna L. Kelly$^2$, Carol Espy-Wilson$^1$ }}

\vspace{-20pt}
\address{\small \em $^1$Univeristy of Maryland College park\\
\small \em $^2$University of Maryland School of Medicine\\
{\small \tt yashish@umd.edu, Ckitchen@jhu.edu, dlkelly@som.umaryland.edu, espy@umd.edu}}

\begin{document}
\vspace{-10pt}
\maketitle

\vspace{-15pt}
\begin{abstract}
This study investigates the speech articulatory coordination in schizophrenia subjects exhibiting strong positive symptoms (e.g. hallucinations and delusions), using a time delay embedded correlation analysis. We show that the schizophrenia subjects with strong positive symptoms and who are markedly ill pose complex coordination patterns in facial and speech gestures than what is observed in healthy subjects. This observation is in contrast to what previous studies have shown in Major Depressive Disorder (MDD), where subjects with MDD show a simpler coordination pattern with respect to healthy controls or subjects in remission. This difference is not surprising given MDD is necessarily accompanied by Psychomotor slowing (i.e., negative symptoms) which affects speech, ideation and motility.  With respect to speech, psychomotor slowing results in slowed speech with more and longer pauses than what occurs in speech from the same speaker when they are in remission and from a healthy subject. Time delay embedded correlation analysis has been used to quantify the differences in coordination patterns of speech articulation. The current study is based on 17 Facial Action Units (FAUs) extracted from video data and 6 Vocal Tract Variables (TVs) obtained from simultaneously recorded audio data. The TVs are extracted using a speech inversion system based on articulatory phonology that maps the acoustic signal to vocal tract variables. The high-level time delay embedded correlation features computed from TVs and FAUs are used to train a stacking ensemble classifier fusing audio and video modalities. The results show that there is a promising distinction between healthy and schizophrenia subjects (with strong positive symptoms) in terms of neuromotor coordination in speech. 

\end{abstract}
\vspace*{1em}
\noindent \textbf{Keywords:} Schizophrenia, Positive symptoms, Facial Action Units, Vocal Tract Variables, Neuromotor coordination

\vspace{-5pt}
\section{Introduction}
Schizophrenia is a chronic mental disorder with heterogeneous presentations that affect around 60 million (1\%) of the world’s adult population (\cite{kuperberg}). Symptoms of schizophrenia are broadly categorized as positive, which are pathological functions not present in healthy individuals (e.g., hallucinations and delusions); negative, which involve the loss of functions or abilities (e.g., apathy, lack of pleasure, blunted affect and poor thinking); and cognitive (deficits in attention, memory and executive functioning) (\cite{Andreason}, \cite{demily}). From previous studies it has been found that individuals suffering from major depressive disorder (MDD) are subjected to neurophysiological changes which often alter motor control and thus affects mechanisms controlling speech production and facial expressions. Clinically these changes are associated with psychomotor slowing,  which is a condition of slowed neuromotor output causing slowed speech, decreased movement and impaired cognitive functions (\cite{Buyukdura}). Previous studies have shown promising results in identifying the severity of depression by using coordination features based on the correlation structure of the movements of various articulators (\cite{Espy-Wilson2019}). This motivated us to investigate how neuromotor coordination is altered in schizophrenic patients who are markedly ill and exhibit strong positive schizophrenic symptoms by analyzing facial activity and speech gestures. 

Previous studies in MDD have used vocal tract variables extracted from audio data (\cite{Seneviratne2020}) and facial action units extracted from video data (\cite{WILLIAMSON2019}) as low level features to classify subjects with MDD from healthy. Time-delay embedded correlation (TDEC) analysis has shown promising results in assessing neuromotor coordination in MDD, and normalized eigenspectra derived from the low level features have been used to develop those classifiers (\cite{WILLIAMSON2019},\cite{Seneviratne2020}, \cite{Williamson2014}). In this study we extend these experiments to assess neuromotor coordination in speech of subjects with strong positive symptoms in schizophrenia. We also show that fusion of audio and video modalities to come up with a multi-modal system results in better classification metrics. 

In Section 2, we explain the dataset, the estimation of the FAUs and TVs, computation of the coordination features, and the details of the classification experiments. Section 3 describes our results in terms of eigenspectra plots and classification outcomes. Interpretation of the results and planned future studies are described in section 4

\vspace{-5pt}
\section{Methods}

\subsection{Dataset Description}
A database recently collected for a collaborative observational study conducted by the University of Maryland School of Medicine and the University of Maryland College Park has been used for this study(\cite{UMBdataset}). The database contains video and audio data of free response assessments administered in an interview format. Data for this study was collected from
23 schizophrenic patients, 18 patients with MDD and 20 healthy controls. All of the schizophrenic and MDD patients were clinically diagnosed. Every subject participated in four interview sessions over a period of six weeks. Each interview session is 10-45 minutes long and every subject is assessed using standard depression severity measures and global psychopathology measures by a clinician and themselves. For this study, we used the clinician assessments based on the 18-item Brief Psychiatric Rating Scale(BPRS), where we selected subjects based on the total BPRS score, and the subscores for psychosis (BPRS item11,item12,item4, item15) and activation (BPRS item6, item7, item17), and the Hamilton Rating Scale for Depression (HAMD). Table \ref{table1:umb_dataset} lists the details of the Dataset. 

Table \ref{table2:our_dataset} presents the information on the subset of data used for our study. The 6 schizophrenic subjects are selected such that they are markedly ill (BPRS total $\geq$ 45), have higher sub-scores for psychosis and activation, but are not depressed or only mildly depressed (HAMD between 0 and 14). The 6 healthy controls are chosen such that they are not depressed
(HAMD < 7) or schizophrenic (BPRS < 32). The MDD subjects are chosen such that they are severely depressed (HAMD $\geq$ 20) but are not schizophrenic (BPRS <32). For this preliminary study we only used data from a single session of the patient's visits. 
\vspace{-5pt}

\begin{table}[th]
\caption{{\it Details on the UMCP-UMB dataset}}
\vspace{-10pt}
\label{table1:umb_dataset}
\footnotesize
\vspace{2mm}
\centering
\begin{tabular}{|c|c|}
\hline
Longitudinal & 5 weeks \\ \hline
Number of Subjects & 31 Male, 30 Female \\ \hline
Demography & 26 African American, 28 Caucasian, 5 Asian \\ \hline
Assessment & HDRS, MADRS, BPRS, CAPE-42 \\ \hline
Recording Type & Video and Audio \\ \hline
Session Length & 10-50 mins \\ \hline
\end{tabular}
\end{table}

\vspace{-10pt}
\begin{table}[th]
\caption{{\it Details on the subset of data used for the study}}
\vspace{-10pt}
\scriptsize
\label{table2:our_dataset}
\vspace{2mm}
\centering
\begin{tabular}{|c|c|c|c|}
\hline
 & SZ & HC & MDD \\ \hline
Number of Subjects & 6 & 6 & 3 \\ \hline
BPRS score range & 45<score<=62 & 18<score<=23 & 18<score<=23 \\ \hline
HAMD score range & 0<score<14 & 0<score<7 & 20<=score \\ \hline
Mean session duration & 35 min & 18 min & 38 min \\ \hline
\end{tabular}
\end{table}

\vspace{-15pt}
\subsection{Vocal Tract Variables (TVs)}
\begin{figure}[th]
\centering
\includegraphics[scale=.15]{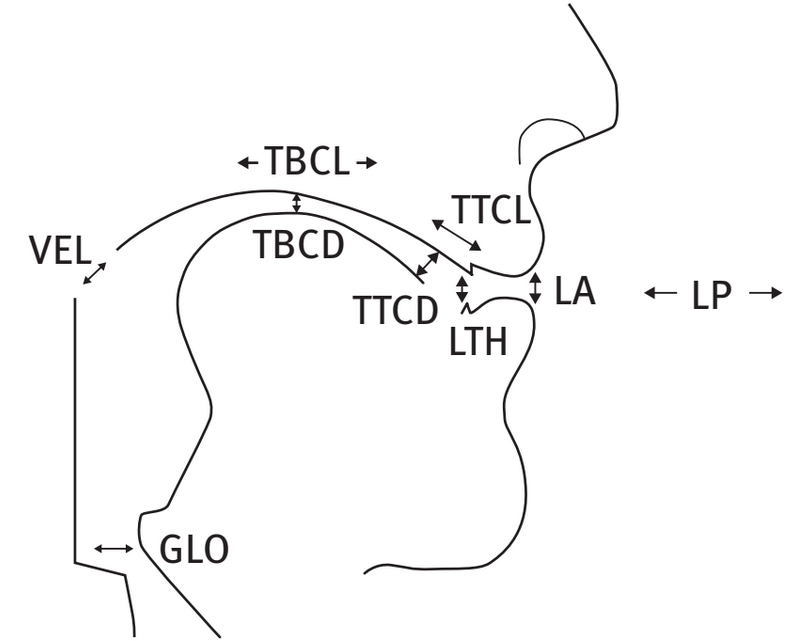}
\caption{Visual representation of the vocal tract variables
at five distinct constriction organs (taken from Saltzman \& Munhall (\cite{Saltzman})), along with a listing of constrictors and their vocal tract variables. See Table 3 for TV labels}  
\label{figure:TVs}
\end{figure}

We used a speech inversion system (\cite{Sivaraman2016}, \cite{Sivaraman2017a}) developed based on Articulatory Phonology (AP) (\cite{Browman1992}) that maps the acoustic signal into vocal tract variables (TVs). The TVs define the kinematic state of each constrictor by its corresponding constriction degree and location coordinates (refer Table \ref{table3:TVs} and Figure \ref{figure:TVs} for more details). The speech inversion systems samples TVs at 100 Hz sampling rate.

\begin{table}[th]
\caption{{\it List of TVs and constrictors}}
\vspace{-10pt}
\label{table3:TVs}
\vspace{2mm}
\footnotesize
\centering
\begin{tabular}{|c|c|}
\hline
\textbf{Constrictors} & \textbf{Vocal Tract Variables (TVs)} \\ \hline
Lip & Lip Aperture (LA), Lip Protrusion(LP) \\ \hline
Tongue Tip & Tongue tip constriction degree (TTCD), \\ \hline
 &Tongue tip constriction location (TTCL) \\ \hline
Tongue Body & Tongue body constriction degree (TBCD), \\ \hline
 &Tongue body constriction location (TBCL) \\ \hline
Velum & Velum (VEL) \\ \hline
Glottis & Glottis (GLO) \\ \hline
\end{tabular}
\end{table}

\vspace{-10pt}

\subsection{Facial Action Units (FAUs)}
The video-based Facial Action Units (FAUs) provide a formalized method for identifying changes in facial expressions. We used the Openface 2.0: Facial Behaviour Analysis toolkit (\cite{openface}) to extract seventeen FAUs (FAU 1,2,4,5,6,7,9,10,12,14,15,17,20,23,25,26 and 45 as in FACS coding system (\cite{Prince2015FacialAC})) from the recorded videos of the subjects during the interviews. The FAU features were sampled at a rate of 28 frames per second. We only analyzed those portions of the video when the subject was talking. The features computed by the tool for the entire video were segmented based on timestamps extracted from manually transcribed transcripts from the audio and relevant speaker ID for the subject. 
\vspace{-5pt}
\subsection{Calculating Coordination Features for Healthy, MDD and Schizophrenic subjects}
Coordination among the seventeen FAUs and among the six TVs (LA, LP, TTCD, TTCL, TBCD and TBCL) were estimated using the correlation structure features. These features are estimated by computing a channel delay correlation matrix using time delay embedding at a fixed delay scale (\cite{Espy-Wilson2019},\cite{WILLIAMSON2019}). For FAUs,  3 samples was chosen as the delay scale and it corresponds to 3/28 = 107 ms and for TVs, 7 samples was chosen as the delay scale and it corresponds to 7/100 = 70 ms. For FAUs, each correlation matrix  has a dimensionality of (255 x 255) with 17 channels and 15 time delays per channel. For TVs, each correlation matrix is (90 x 90) dimensional with 6 channels and 15 time delays per channel. After speech diarization, to calculate correlation features, only the segments of the subject which are greater than 5 seconds were used. 

From the correlation matrix $R_{i}$ calculated for each sample $i$, the eigenspectrum is computed. The eigenspectrum generated for FAUs is a 255- dimensional vector which is rank ordered (in the descending order of magnitude of eigenvalues) from index j=1,..,255. The eigenspectrum generated from TVs is a 90-dimensional vector rank ordered from index j=1,..,90. 

The eigenspectrum generated can be considered as a high level feature designed to characterize properties of coordination and timing from the low level features (\cite{WILLIAMSON2019}). The eigenspectrum characterizes the within-channel and cross-channel distributional properties of the multivariate FAU and TV time series. The magnitude of the eigenvalues represent the average correlation in the direction of corresponding eigenvectors. Therefore the significance of the magnitude of eigenvalues indicate the number of independent dimensions that can be used to represent speech belonging to different groups. Therefore, a few significant eigenvalues imply a simpler articulatory coordination pattern whereas a large number of significant eigenvalues correspond to more complex articulatory coordination. 

\vspace{-8pt}
\subsection{Classification Between Schizophrenia and Healthy Subjects}
\label{sec:classification}

From Table \ref{table2:our_dataset}, all the 6 Schizophrenic subjects and all the 6 Healthy controls are chosen to train a Support Vector Machine (SVM) classifier with radial basis function kernel. The classifier was trained on the coordination features computed over FAUs and TVs to classify a given subject as a schizophrenic subject or a healthy control. Eigenvalues were averaged over multiple index ranges of the normalized eigenspectrum (equations used for calculating normalized eigenspectra are from (\cite{Espy-Wilson2019})) to be used as the input features to the classifier. The features calculated are standardized across all instances before model training and testing. We first trained individual SVM models for TVs and FAUs from eigenspectra features and then trained a fused model by combining TV and FAU features using a stacking ensemble model(\cite{WOLPERT1992241}). 

The SVM models were trained and evaluated in leave-one-subject-out-cross-validation fashion with a total of 12 folds. The average accuracy and F1 scores are computed across all folds.

\vspace{-5pt}
\section{Results}
\begin{figure}[t]
  \centering
  \begin{subfigure}{1.0\columnwidth}
    \includegraphics[width=1.0\linewidth,height=45mm]{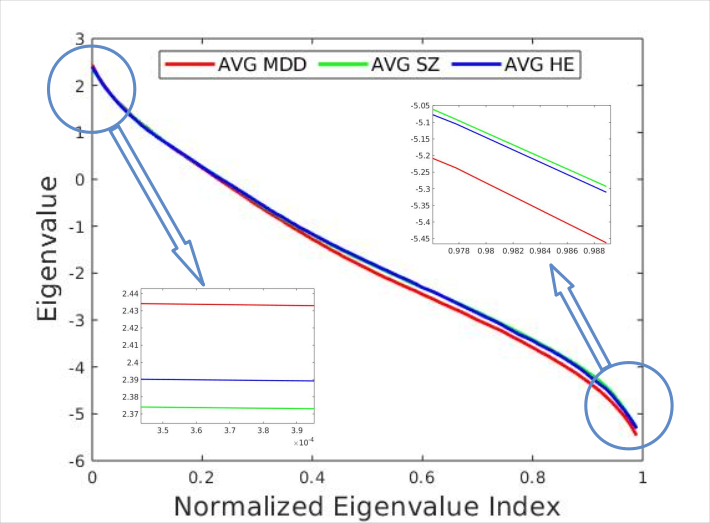}
    \subcaption{Averaged eigenspectra from TVs}
  \end{subfigure}
  \begin{subfigure}{1.0\columnwidth}
    \includegraphics[width=1.0\linewidth,height=45mm]{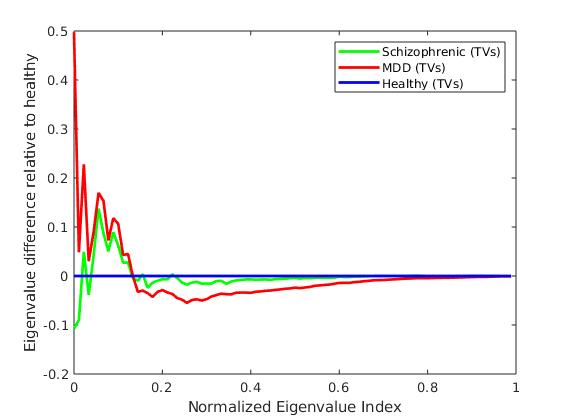}
    \subcaption{Difference plot}
  \end{subfigure}
  \caption{ Averaged eigenspectrum for TVs (left) and corresponding difference plot (right)}
  \label{fig2:avg_eigenspectra_tvs}
\end{figure}

\begin{figure}
\begin{multicols}{2}
    \includegraphics[width=1.15\linewidth]{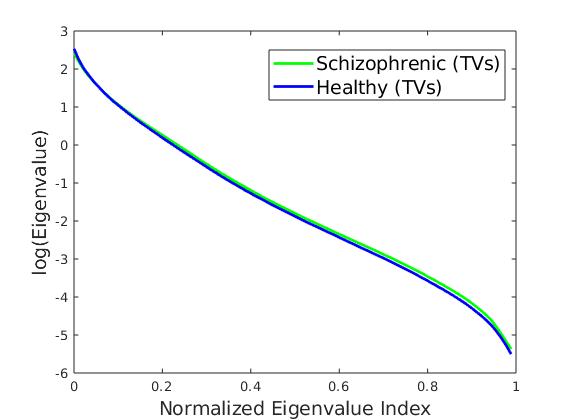}\par 
    \includegraphics[width=1.15\linewidth]{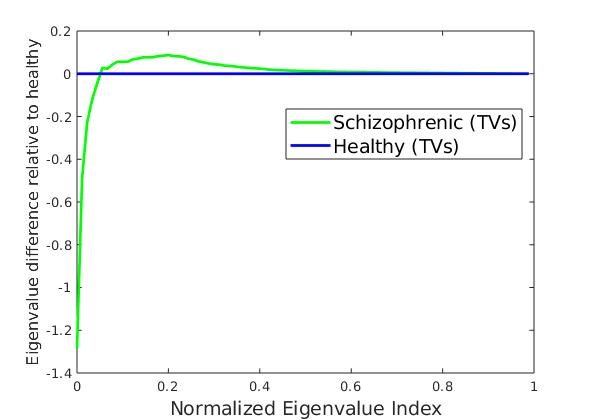}\par 
    \end{multicols}
\begin{multicols}{2}
    \includegraphics[width=1.15\linewidth]{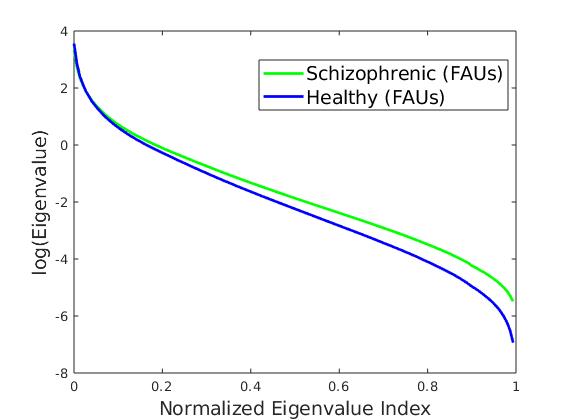}\par
    \includegraphics[width=1.14\linewidth]{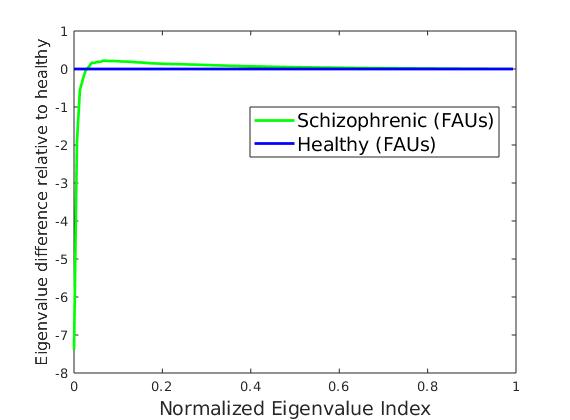}\par
\end{multicols}
\caption{ Averaged eigenspectra for TVs and FAUs (left) and corresponding difference plots (right) for classification experiments}
\label{fig3:eigenspectra_faus_tvs}
\end{figure}

In the first experiment, we compared 3 subjects from each schizophrenic, MDD and healthy groups by calculating the eigenspectra from TVs and FAUs and their corresponding difference plots. Figure \ref{fig2:avg_eigenspectra_tvs} shows the averaged eigenspectra plot and the corresponding difference plot obtained from TVs. The eigenvalues are plotted in the logarithmic scale and the plot is zoomed in at low and high rank indices to see where the curves lie with respect to each other. The difference curves for schizophrenia and MDD in the difference plot are calculated relative to healthy.

Figure \ref{fig3:eigenspectra_faus_tvs} shows the eigenspectra and difference plots obtained from both TVs an FAUs for the classification experiment. It also confirms the agreement between coordination patterns seen in TVs and FAUs for schizophrenia subjects.

Table \ref{table4:classifciation} shows the average accuracies and F1 scores obtained from the classification experiments in section \ref{sec:classification}. The highest accuracy of 68.19\% (F1 scores of 70.12 for schizophrenic group and 65.23 for healthy group) was achieved from the fused model which is a promising improvement with respect to the individual modalities. 

\vspace{-5pt}

\begin{table}[th]
\caption{{\it Classification Results}}
\vspace{-10pt}
\footnotesize
\label{table4:classifciation}
\vspace{2mm}
\centering
\begin{tabular}{|c|c|c|c|}
\hline
Method & Index range & Accuracy & F1(S)/F1(H) \\ \hline
FAU & [0-0.02],[0.96-1] & 65.63 \% & 67.89/61.37\\ \hline
TV & [0-0.03],[0.95-1] & 61.68 \% & 63.45/59.21 \\ \hline
\textbf{Multi-modal} & - & \textbf{68.19 \%} & \textbf{70.12/65.23} \\ \hline
\end{tabular}
\end{table}

\vspace{-15pt}
\section{Discussion}
Figure \ref{fig2:avg_eigenspectra_tvs} shows that the low rank eigenvalues are larger for MDD subjects  relative to the schizophrenic patients and the healthy controls, and this trend is reversed towards the high rank eigenvalues. This pattern is a key observation associated with depression severity (\cite{Williamson2014},\cite{WILLIAMSON2019},\cite{Espy-Wilson2019}). The magnitude of high rank eigenvalues indicates the dimensionality of the time-delay embedded feature space. Thus, larger values in the high rank eigenvalues can be associated with greater complexity of articulatory coordination (\cite{Espy-Wilson2019}). Thus, we can conclude that the schizophrenic subjects with strong positive symptoms have a higher articulatory coordination complexity than the healthy controls and the MDD patients, and the MDD patients have a simpler articulatory coordination pattern relative to the healthy controls and the schizophrenic patients.  These results are likely due to the negative symptoms of depression which results in psychomotor slowing (i.e., simpler coordination) and the strong positive symptoms of the schizophrenic patients such as activation that results in motor hyperactivity (i.e., complex coordination).  We see this effect in both the eigenvalues computed from the FAUs and from the TVs.

The classification results indicate that there is a notable discrimination between the coordination features for schizophrenic subjects with strong positive symptoms and those of healthy subjects.  From this preliminary study, we observe that facial gestures were more effective compared to TVs in the classification experiments. This could be because of the inclusion of a wider range of facial muscle movements which were not limited to only those around the speech articulators. From previous studies it has been shown that some FAUs are significant in understanding depression severity (\cite{Girard}). Following that line, we could come up with attention based deep learning models to select the most discriminative set of FAUs, from which the performance of the classification models can be further improved.
Finally, as Seneviratne et al. (\cite{Seneviratne2020}) have shown, the performance of the TV based classification models can be improved by adding glottal TVs to the constriction degree and location TVs in detecting subjects with severe depression. We will investigate the use of these glottal TVs as well as the velar TVs to get a full representation of the speech gestures and their coordination.   

It should also be noted that (Tron et al.\cite{Tron}) found a strong correlation between negative symptoms of schizophrenia (e.g., blunted affect) and various facial dynamics. Further, there have been other studies(\cite{Tremeau}) where the schizophrenic and the depressed patients were compared with healthy controls using facial expressiveness in terms of negative symptoms. The study of Tremeau et al.(\cite{Tremeau}) observed similar deficits in both the depressed and schizophrenic subjects. But our study, which focused on differentiating subjects with strong positive symptoms based on coordination features, presents the first evidence that the positive symptoms of schizophrenia can be characterized by complex articulatory coordination pattern of the speech and facial gestures.  

In future work, we plan to validate these preliminary findings using a larger dataset. We are also working on developing a Multi-modal Convolutional Neural Network (CNN) based deep learning model where the correlation matrices are fed directly to perform classification.   

\vspace{-8pt}
\section{Acknowledgements}
This work was supported by a UMCP UMB - AI + Medicine for High Impact (AIM-HI) Challenge Award. We would like to thank our AIM-HI group for valuable discussions and providing the transcripts of the clinical interviews

\vspace{-5pt}
\eightpt
\printbibliography[heading=refs]
\end{document}